\documentclass[12pt,preprint]{aastex}

\usepackage{apjfonts} 

\begin{document}

\title{Cosmology from very high energy $\gamma$-rays}

\author{Xiao-Jun Bi\altaffilmark{1,2} and Qiang Yuan\altaffilmark{2}}

\altaffiltext{1}{Center for High Energy Physics, Peking University, Beijing 
100871, P.R.China}
\altaffiltext{2}{Key Laboratory of Particle Astrophysics, Institute of High 
Energy Physics, Chinese Academy of Sciences, Beijing 100049, P.R.China
}

\begin{abstract}

In this work we study how the cosmological parameter, the Hubble 
constant $H_0$, can be constrained by observation of very high
energy (VHE) $\gamma$-rays at the TeV scale. The VHE $\gamma$-rays 
experience attenuation by background radiation field through 
$e^+e^-$ pair production during the propagation in the intergalactic 
space. This effect is proportional to the distance that the VHE 
$\gamma$-rays go through. Therefore the absorption of TeV 
$\gamma$-rays can be taken as cosmological distance indicator to 
constrain the cosmological parameters. Two blazars Mrk 501 and 1ES 
1101-232, which have relatively good spectra measurements by the 
atmospheric Cerenkov telescope, are studied to measure $H_0$.
The mechanism measuring the Hubble constant adopted here is very 
different from the previous methods such as the observations of 
type Ia supernovae and the cosmic microwave background. However, at
$2\sigma$ level, our result is consistent with which given by other 
methods.

\end{abstract}


\section{Introduction}

The modern cosmology achieves great progress in recent years. A 
concordance $\Lambda$CDM cosmology has been built thanks to the 
precise observations of the ``distance indicators'' type Ia supernovae 
\citep[SNe Ia,][]{1998AJ....116.1009R,2004ApJ...607..665R,
1999ApJ...517..565P} and the anisotropy of cosmic microwave 
background \citep[CMB,][]{2000Natur.404..955D,2003ApJS..148..175S}. 
There are also other cosmological probes such as the large scale structures 
\citep{2004ApJ...606..702T}, galaxy clusters \citep{2008MNRAS.383..879A}, 
observational Hubble parameters \citep{2005PhRvD..71l3001S} and the weak 
gravitational lensing \citep{2008PhR...462...67M} further supporting this 
scenario. Different methods are roughly consistent with each other within 
the observation uncertainties. It is very important to develop additional
complementary observational evidence to test this model and measure the
cosmological parameters.

It has been pointed out that the observations of VHE $\gamma$-rays at 
the energy scale of hundred GeV to TeV scale are possible to provide 
another independent constraint on the cosmological parameters
\citep{1994ApJ...423L...1S}.
Thanks to the rapid technical development of VHE $\gamma$-ray detection,
especially the atmospheric Cerenkov telescopes, great progress of VHE
$\gamma$-ray astronomy is achieved in recent years and a large number 
of VHE $\gamma$-ray sources are detected. Even $\gamma$-ray sources at 
cosmological distances, such as Mrk 501 at $z=0.034$ and 1ES 1101-232 at 
$z=0.186$ are observed. More importantly the spectra of these sources 
have been measured with relatively high precision, which provide us the 
possibility to untangle the effect of attenuation when $\gamma$-rays 
propagate in the intergalactic space.

The attenuation of VHE $\gamma$-rays is induced by the electron-positron 
pair production $\gamma+\gamma_{bk} \rightarrow e^+ + e^-$ during its 
propagation in the background radiation field \citep{Nikishov1962,
1967PhRv..155.1404G,1966PhRvL..16..252G}. This process is actually 
complex. The observed spectra of extragalactic sources are related 
with several issues: the intrinsic 
spectra at sources, the cross section of $\gamma\gamma$ interaction, 
the intensity of the cosmic infrared background (CIB), and the physical 
distance the VHE $\gamma$-rays cross. Even before the first detection 
of the VHE $\gamma$-rays from distant extragalactic sources, the 
perspective to explore the CIB using the attenuation effect was proposed 
\citep{1992ApJ...390L..49S}. The discovery of the first extragalactic VHE 
$\gamma$-ray source, an active galactic nuclei (AGN) Mrk 421, was 
performed by Whipple in 1992 \citep{1992Natur.358..477P}. Till now more 
than 20 extragalactic sources, most of which are AGNs, are discovered by 
ground-based observatories\footnote{See the VHE source web by Wagner, 
http://www.mppmu.mpg.de/$\sim$rwagner/sources/}. The observations of these 
sources provide us valuable information in understanding the $\gamma$-ray 
production mechanism and give useful implication or constraint on the CIB 
intensity \citep[e.g.,][]{1999APh....11...35C,2000A&A...353...97K,
2001A&A...371..771R,2006Natur.440.1018A}. On the other hand, once the
primary spectra and the CIB intensity are specified, the distance-redshift
relation (accordingly the cosmological model parameters) of the sources 
can be derived from the absorption effect. 

In this work we try to constrain the Hubble constant from the absorption
effect of distant  VHE $\gamma$ sources by the CIB. By a global fitting 
to the observational spectra of two TeV blazars, Mrk 501 and 
1ES 1101-232, we get the Hubble constant with larger errors compared with
other methods. 
In our work the $\Lambda$CDM universe with matter component 
$\Omega_M=0.28$ and dark energy $\Omega_{\Lambda}=0.72$ is adopted 
\citep{2008arXiv0803.0547K}. We find that the best-fitting to the data 
of the two sources intend to give similar Hubble 
constant, although they have very different intrinsic spectra and 
redshifts. This is very encouraging that the attenuation may indeed 
give implications on the cosmological parameters.
We noticed in a previous work \cite{2008arXiv0804.3699B} adopt
the similar effect to derive the lower limit 
of the Hubble constant from observation of Mrk 501.
In their work, the direct measurements of 
CIB intensity was adopted 
and the intrinsic spectrum of the source was required to be concave. 

The outline of this paper is as follows. Sec. 2 describes the absorption
of VHE $\gamma$ photons by CIB. In Sec. 3 we present an introduction to
the observations of the two TeV blazars. The implication on Hubble 
constant is given in Sec. 4. Finally we give conclusion and some 
discussion in Sec. 5.

\section{Absorption of TeV $\gamma$-rays in CIB}

The fundamental process of the VHE $\gamma$-ray absorption is due to
electron/positron pair production $\gamma+\gamma_{bk} \rightarrow 
e^+ + e^-$. The threshold energy of the pair production is 
$m_e^2/\epsilon$, with $\epsilon$ the energy 
of the background radiation. For the CMB photon $\epsilon\sim 10^{-3}$ 
eV, this absorption takes place for $\gamma$-rays with energy 
$E\gtrsim 1$ PeV. While for the TeV scale $\gamma$-rays that the current 
experiments can probe, the responsible soft photon is in the infrared
band, i.e., CIB with $\epsilon\sim 1$ eV ($\lambda\sim 1$ $\mu$m). An
approximate relation between energies of attenuated VHE $\gamma$-rays 
and the CIB photons is
\begin{equation}
\frac{\lambda}{1\mu {\rm m}}\sim 1.2 \frac{E}{\rm 1TeV}\ .
\end{equation}

The observed VHE $\gamma$-ray spectrum after attenuation is given by 
\begin{equation}
F_{\rm obs} = e^{-\tau} F_{\rm int}\ , 
\end{equation}
where $\tau$ is the optical depth and $F_{\rm int}$ is the intrinsic 
spectrum at the source. For the CIB with number density $n(\epsilon)$, 
the optical depth $\tau$ is given as \citep{1967PhRv..155.1404G}
\begin{equation}
\tau(E)=\int {\rm d}l\int {\rm d}\cos\theta \frac{1-\cos\theta}{2}
\int {\rm d}\epsilon n(\epsilon)\sigma(E,\epsilon,\cos\theta),
\label{tau}
\end{equation}
where ${\rm d}l=c{\rm d}t=\frac{c}{H_0}\frac{{\rm d}z}{(1+z)
\sqrt{0.28(1+z)^3+0.72}}$ is the differential path traversed by the VHE 
$\gamma$-rays,
$\theta$ is the angle between the momenta of VHE $\gamma$-ray and CIB
photon. The cross section of pair production is
\begin{eqnarray}
\sigma(E,\epsilon,\cos\theta)& = & \sigma_T\cdot\frac{3m_e^2}{2s}\cdot\left[
-\frac{p_e}{E_e}\left(1+\frac{4m_e^2}{s}\right)+\right. \nonumber \\
& & \left.\left(1+\frac{4m_e^2}{s}
\left(1-\frac{2m_e^2}{s}\right)\right)\log\frac{(E_e+p_e)^2}{m_e^2}\right],
\label{sigma}
\end{eqnarray}
with $\sigma_T=6.65\times 10^{-25}$ cm$^2$ the Thomson cross section, 
$s=2E\epsilon(1-\cos\theta)(1+z)^2$ the center of momentum system (CMS) 
energy square, $E_e=\sqrt{s}/2$ and $p_e=\sqrt{E_e^2-m_e^2}$ the CMS 
energy and momentum of electrons.

\begin{figure}[!htb]
\centering
\includegraphics[width=\columnwidth]{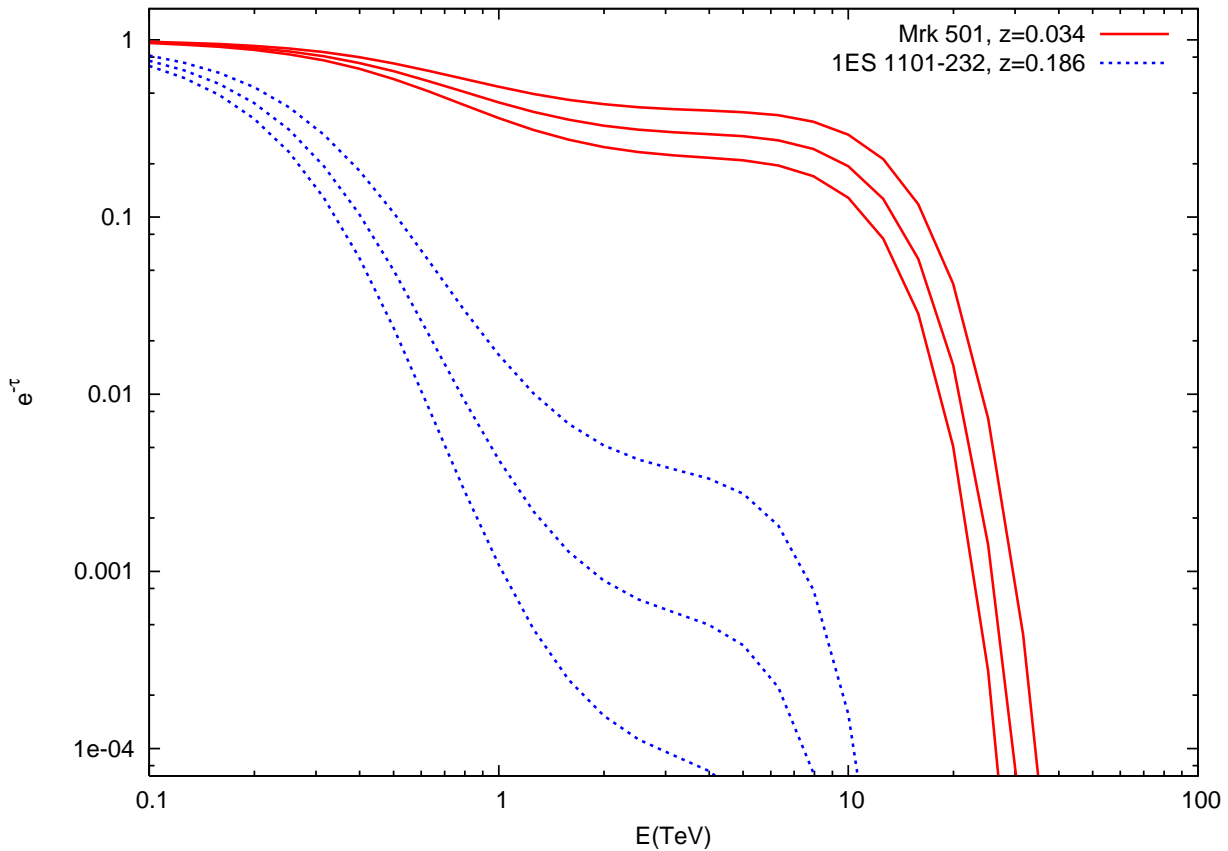}
\caption {Attenuation factor of  VHE $\gamma$-rays for sources Mrk 501 
and 1ES 1101-232. The three curves for each source correspond to the
``nominal'' (middle curve), $25\%$ higher (lower curve) and $25\%$ 
lower (upper curve) CIB respectively. The Hubble constant in the 
calculation is adopted as $h=0.7$.}
\label{attenuation}
\end{figure}

From Eq. (\ref{tau}) we can see that the intensity of CIB is crucial in
determining the effect of attenuation. The CIB is generated by stars and 
absorption/re-emission of star light by dust in galaxies. The status of 
measurements and models of CIB can be found in the review paper by 
\cite{2001ARA&A..39..249H}. Because of the contamination of foreground 
from the solar system and the Galaxy, the determination of CIB has 
relative large uncertainty. Here we adopt the ``nominal'' model prediction
of \cite{2001ICRC...27I.250A} (curve 1 of Fig. 1) which can give a good 
description of the measurements. Two other models (curves 2 and 3 in 
Fig. 1 of \cite{2001ICRC...27I.250A}) are regarded as the lower and 
upper limits of CIB intensity. The differences between these model
predictions can vary from several tens percent to several times at
different energies. To simplify the uncertainties of CIB in our analysis 
we take the uncertainty of $\pm 25\%$ relative to the ``nominal'' model 
of \cite{2001ICRC...27I.250A} to represent the upper and lower limits.
The CIB is denoted as $n(\epsilon)=A\bar{n}(\epsilon)$, where 
$\bar{n}(\epsilon)$ represents the best \citep{2001ICRC...27I.250A} model 
of CIB, $A=1\pm 25\%$ is a normalization factor to represent the 
uncertainties, which is energy independent. This form of uncertainties 
greatly simplifies the process of global fitting.

The comoving density of CIB is adopted to be constant without redshift 
evolution, which is shown to be of little influence for the sources with 
redshift $z\lesssim 0.2$ \citep{2006Natur.440.1018A}. Using this CIB 
field, we calculate the attenuation factor $e^{-\tau}$ of VHE 
$\gamma$-rays according to Eq.(\ref{tau}) for sources Mrk 501 ($z=0.034$) 
and 1ES 1101-232 ($z=0.186$), as shown in Fig. \ref{attenuation}. 
It can be seen from this figure that the absorption for $\gamma$-rays 
increases rapidly for energies $\gtrsim 10$ TeV. It also shows 
that absorption of the nearby source Mrk 501 at energies $\sim 20$ TeV 
is comparable with the effect to the distant source 1ES 1101-232 at 
energies $\sim$TeV \citep{2008arXiv0804.3699B}. Therefore $\gamma$-rays 
with high energy ($E\gtrsim 10$ TeV) or high redshift ($z\gtrsim 0.1$) 
will be very effective to study the attenuation process and CIB 
\citep{1999APh....11...93P}.

\section{TeV $\gamma$-ray Observations of the sources: Mrk 501 and 
1ES 1101-232}

\begin{figure*}[!htb]
\centering
\includegraphics[width=0.45\columnwidth]{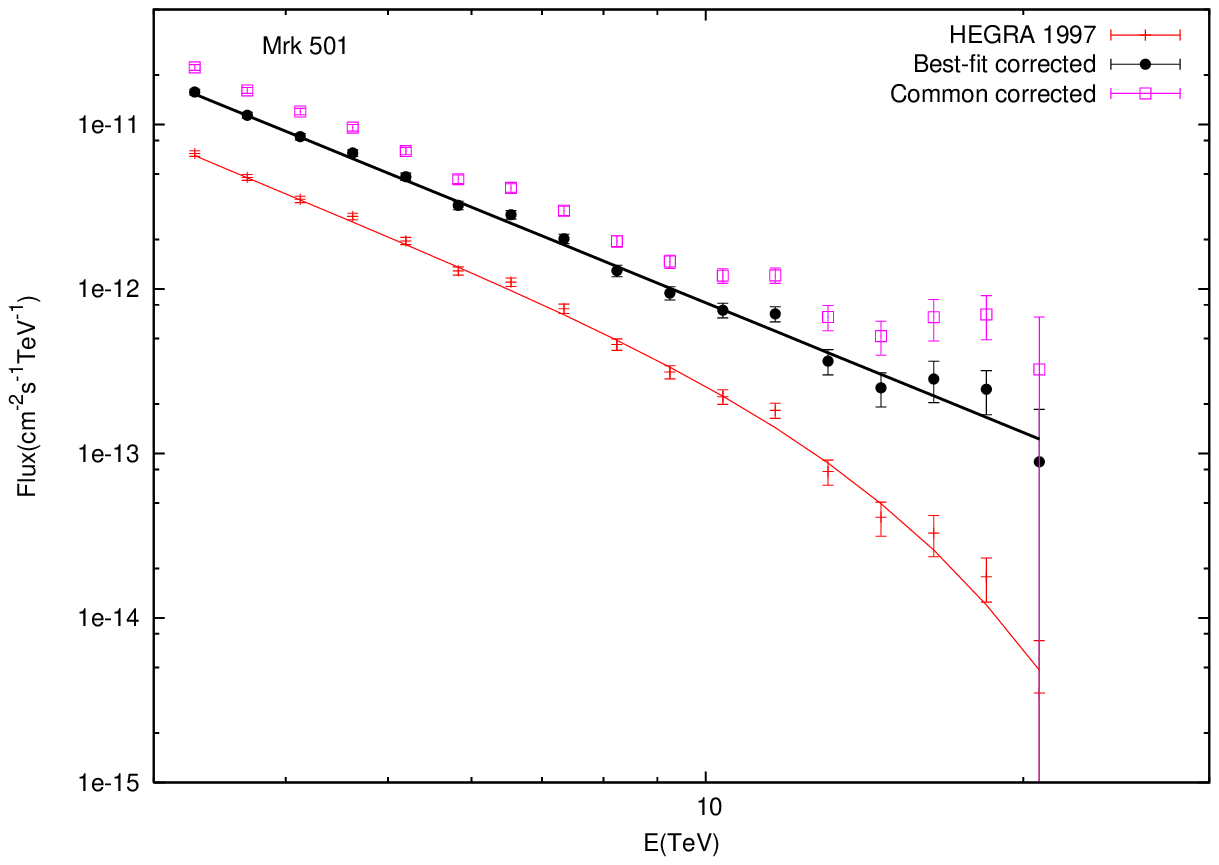}
\includegraphics[width=0.45\columnwidth]{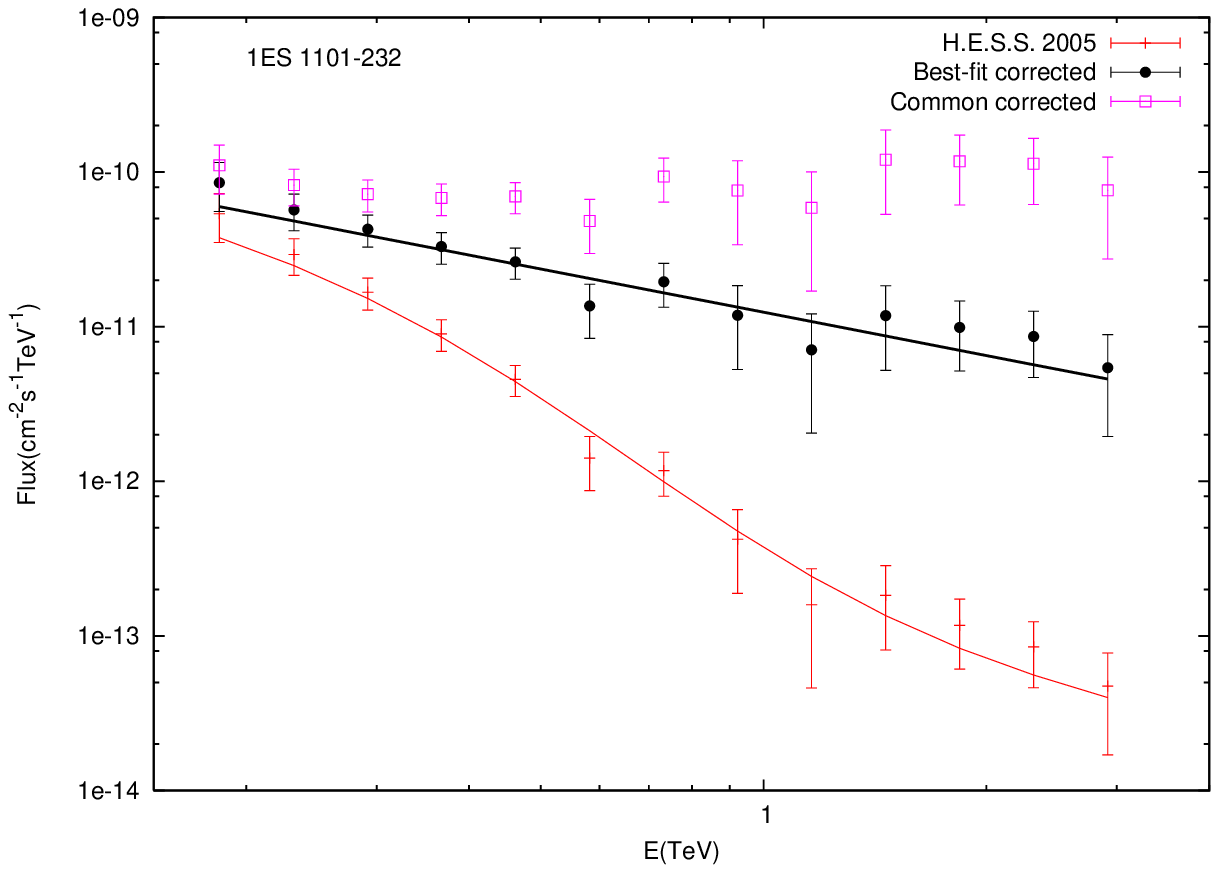}
\caption {Left: Energy spectrum of Mrk 501 by HEGRA in 1997 
\citep{2001A&A...366...62A}. Also shown are the absorption-corrected 
spectra using the usually adopted Hubble constant $h=0.7$ (upper
most points) and the best-fitting Hubble constant $h=0.98$ (medium points, 
see below \S 4). The two lines are the best-fitting power law intrinsic 
spectrum (thick-black, $\Gamma_{\rm int}=2.62$) and the absorbed one 
(thin-red). Right: same as the left panel but for 1ES 1101-232 by H.E.S.S. 
in 2005 \citep{2006Natur.440.1018A}. The best-fitting parameters are 
$h=1.09$ and $\Gamma_{\rm int}=0.93$ instead.}
\label{spectrum} 
\end{figure*}

Mrk 501 is a nearby (with redshift $z=0.034$) BL Lac type blazar, which
is a kind of radio-loud AGN with relativistic jet being aligned along the 
line of sight. The first detection of VHE $\gamma$-ray emission from Mrk 
501 was performed by Whipple in 1995 \citep{1996ApJ...456L..83Q}. During
the 1997 flares, Mrk 501 was observed by several experiment groups 
\citep{1997ApJ...487L.143C,1998ApJ...501L..17S,1997A&A...327L...5A,
1999A&A...350...17D,1998ApJ...504L..71H}. The energy up to $\sim 20$ 
TeV observed from Mrk 501 makes it a good candidate 
to study the absorption effect of high energy $\gamma$-rays in CIB 
\citep{1999APh....11...35C} and the possible Lorentz violation effect 
\citep{2000PhLB..493....1P}. In the current study,
we use the reanalyzed HEGRA data in 1997 with improved energy resolution
\citep{2001A&A...366...62A}. The observational spectrum is shown in the 
left panel of Fig. \ref{spectrum}. 

The other source investigated here, 1ES 1101-232, is a distant blazar with
redshift $z=0.186$. TeV observation of 1ES 1101-232 was performed
by H.E.S.S. in 2005 \citep{2006Natur.440.1018A}. It is shown that even though
the measured maximum energy only reaches $\sim 3$ TeV, 1ES 1101-232 is still
very effective to probe the intensity of CIB in the propagation path VHE 
$\gamma$-rays go through due to its large distance from us 
\citep{2006Natur.440.1018A}. The observational spectrum of 1ES 1101-232 
is shown in the right panel of Fig. \ref{spectrum}. 

Actually these observations show harder spectra than expected.
Considering the effects of absorption of the VHE $\gamma$-rays from
the normal CIB density and distance-redshift relation, the observed 
spectra means unnaturally hard intrinsic spectra at the sources. It can
be clearly seen from the upper most points in Fig. \ref{spectrum},
which are calculated using the usually adopted CIB and $h=0.7$. For
1ES 1101-232, the corrected spectrum is $\sim E^{-0.1}$, which seems
extremely hard when comparing with the expected one from shock
acceleration \citep{2001RPPh...64..429M}. The corrected spectrum of 
Mrk 501 can not be fitted with a single power law, however, it is 
also shown in Fig. \ref{spectrum} that the spectrum of high energy part 
is very flat. 

This possible anomaly has led to quite a few 
discussions about possible new physics. It was proposed that the 
axion-$\gamma$ oscillation when the VHE $\gamma$-rays propagate in the 
intergalactic magnetic field makes the universe more transparent 
than naively expected \citep{2008PhRvD..77f3001S}. When $\gamma$-rays 
oscillate into axions they will not be absorbed by the CIB and keep the
primary spectra unchanged \citep{2007PhRvL..99w1102H}. It was also 
suggested that the possible Lorentz violation may be responsible for 
the hard $\gamma$-ray spectra \citep{2000PhLB..493....1P}. In this
scenario the threshold energy of the interaction moves to higher energy
and the absorption effect at the observed energy scales becomes weaker. 
Possible explanations of the hard spectra within astrophysics are also 
discussed \citep{2006Natur.440.1018A,2008MNRAS.387.1206A}. In
\cite{2006Natur.440.1018A} the authors pointed out that if the CIB 
intensity is about half of the locally measured values the observed 
$\gamma$-ray spectra can be naturally explained. \cite{2008MNRAS.387.1206A}
also suggested some special mechanism to produce very hard intrinsic 
spectra at the sources.

Since all these explanations are based on a standard cosmological model,
we are considering that if these observations have implications on the
cosmological model itself. In the following we will show that the 
best-fitting to the $\gamma$-ray data favors a larger value of the 
Hubble constant. In spite of the large uncertainties, our result is 
consistent with previous cosmological measurements at 2$\sigma$ level.

\section{Implication on the Hubble constant}

The $\gamma$-ray spectra from astrophysical sources are usually very
well described by power law functions, which originated from the shock 
wave acceleration at the sources. Assuming the power law spectral index 
$\Gamma_{\rm int}$ at source we get the observed spectrum 
$F_{\rm obs}\propto E^{-\Gamma_{\rm int}}\,e^{-A\tau_0/h}$ according to
Eq. (\ref{tau}), where $\tau_0$ is the optical depth with $A=1$ and $h=1$. 
Using the observational data we can fit the parameters $\Gamma_{\rm int}$, 
$A$ and $h$. It should be noted that parameters $A$ and $h$ are strongly 
coupled with each other, so it is unable to determined them simultaneously 
from the attenuation of VHE $\gamma$-ray spectra. We firstly fix $A=1$, 
and fit the parameters $\Gamma_{\rm int}$ and Hubble constant $h$. 
Then we will take the uncertainty of $A$ into account. 

\begin{figure*}[!htb]
\centering
\includegraphics[width=0.45\columnwidth]{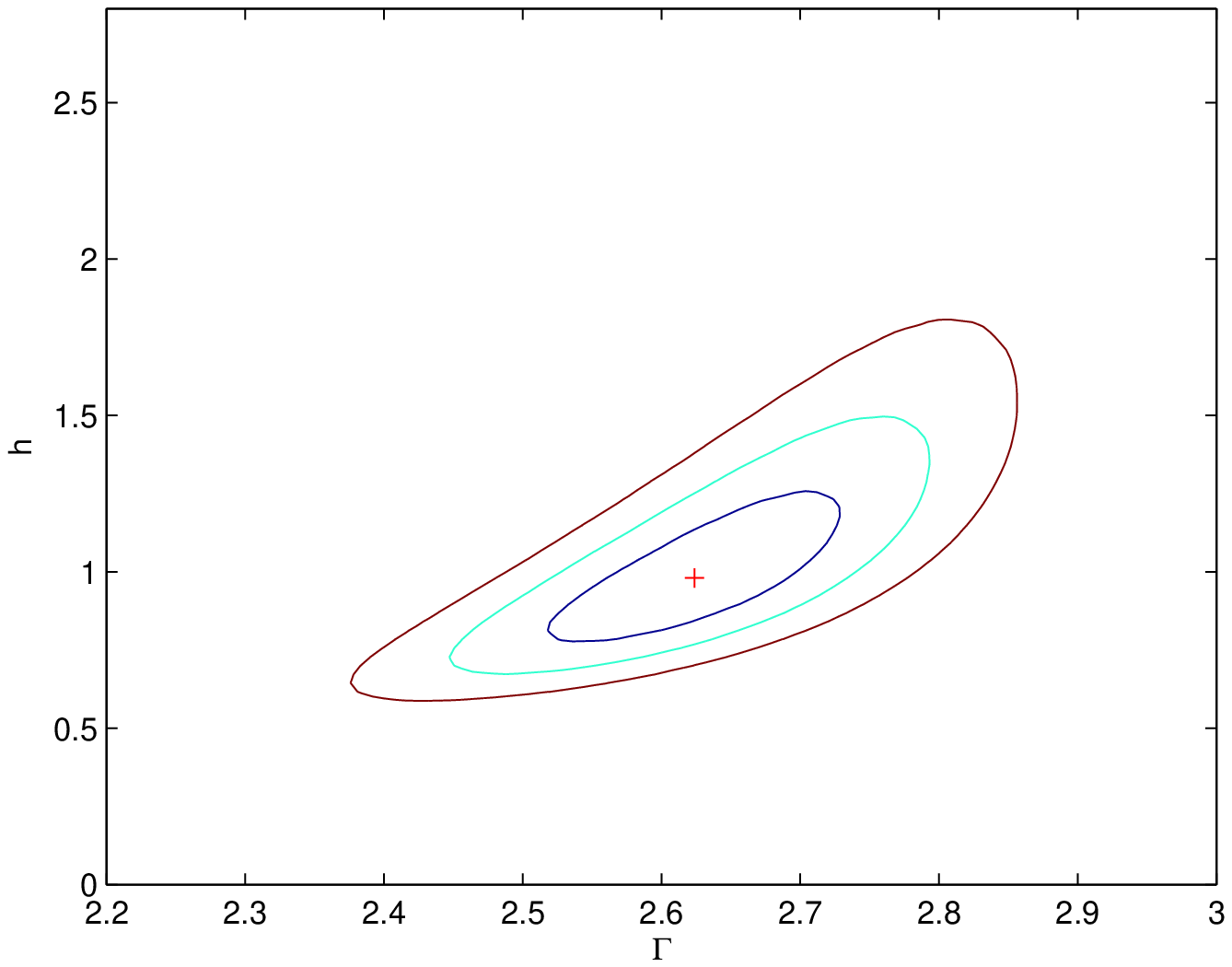}
\includegraphics[width=0.45\columnwidth]{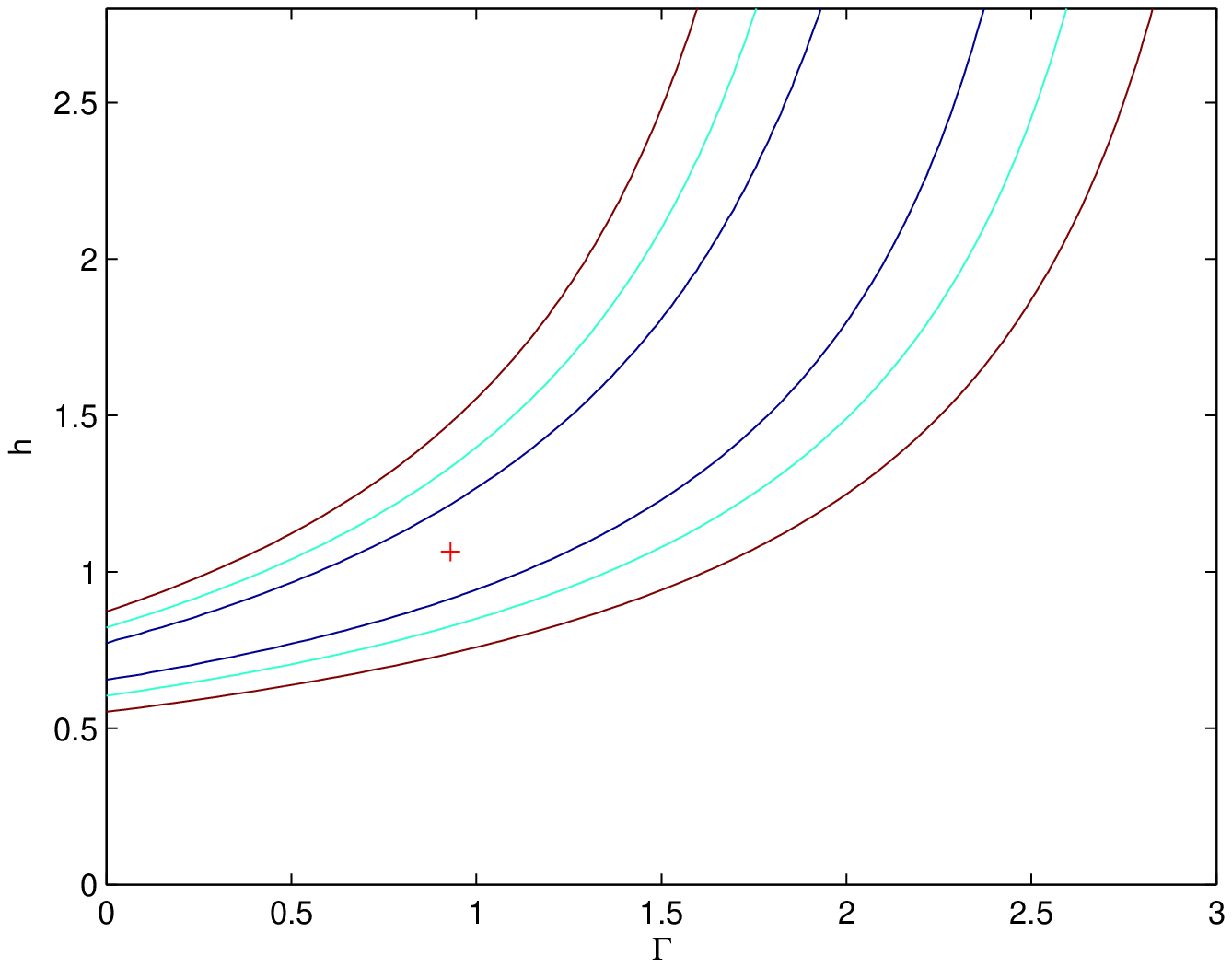}
\caption {$1$, $2$ and $3\sigma$ confidence regions (from inner to outer)
of parameters $\Gamma_{\rm int}$ and $h$ for $A=1$. Left: for Mrk 501; right: 
for 1ES 1101-232. The cross in each panel is the best-fitting value.}
\label{fixA} 
\end{figure*}

The confidence regions in the $h - \Gamma$ plane are shown in Fig. 
\ref{fixA}. The best-fitting values and $1\sigma$ errors of the parameters 
are compiled in Table \ref{fitresult}. It is shown that for both sources, 
the best-fitting Hubble constant $h$ is close to $1$, which is larger than 
the results from other cosmological measurements $h\approx 0.7$, such as 
from SNe Ia \citep{2006A&A...447...31A} and CMB anisotropy 
\citep{2008arXiv0803.0547K}. A larger $h$ implies that smaller absorption 
is favored by the observations. Similar results are also found by the 
previous studies \citep[e.g.,][]{2000PhLB..493....1P,2006Natur.440.1018A}.
We also notice in Fig. \ref{fixA} that since the statistical 
errors of Mrk 501 are much smaller than that of 1ES 1101-232 it also
gives much better constraints on the parameters.

The best-fitting intrinsic spectrum is $\Gamma_{\rm int}=2.62$ for Mrk 
501 and $0.93$ for 1ES 1101-232. It seems that the intrinsic spectrum 
for 1ES 1101-232 is still too hard. Generally the intrinsic spectrum 
of blazars for both hadronic and leptonic scenarios 
from shock acceleration is expected to be $\Gamma_{\rm int}\gtrsim 1.5$
\citep{2001RPPh...64..429M,2006Natur.440.1018A}. If we apply a limit
$\Gamma_{\rm int}\gtrsim 1.5$, we find that the Hubble constant $h>1.2$ 
at $68\%$ confidence level from 1ES 1101-232. It should be noted that
scenario with very hard $\gamma$-ray spectrum is also proposed recently
\citep{2008MNRAS.387.1206A}.

It is also shown in Fig. \ref{fixA} that there is degeneracy between 
the parameters $\Gamma_{\rm int}$ and $h$, especially for the source
1ES 1101-232. This is because a harder $\Gamma_{\rm int}$ means a 
stronger absorption, and leads to a smaller $h$ (or a larger $A$).

The reconstructed source spectra using the best-fitting parameters are
shown by the medium points in Fig. \ref{spectrum}. The thick lines in this
figure represent the best-fitting power law intrinsic spectra. We can 
see that the reconstructed source spectra are well consistent with 
power law functions.

\begin{deluxetable}{lcccc}
\tablecolumns{3} \tablewidth{0pc} \tablecaption{Fitting results for
parameters} \tablehead{Source  & \multicolumn{2}{c}{$A=1$} & \multicolumn{2}{c}
{$A=1\pm0.25$} }\startdata
      & $\Gamma_{\rm int}$ & $h$ & $\Gamma_{\rm int}$ & $h$ \\
 Mrk 501 & $2.62^{+0.11}_{-0.10}$ & $0.98^{+0.28}_{-0.20}$ & $2.62^{+0.12}_{-0.10}$ & $1.01^{+0.53}_{-0.40}$ \\
 1ES 1101-232\tablenotemark{a} & 0.93 & 1.09 & 2.15 & 2.80 \\
 Combined & --- & $1.00^{+0.15}_{-0.14}$ & --- & $1.05^{+0.35}_{-0.19}$
\enddata
\tablenotetext{a} {The fitting errors of parameters for this source
are very large that not shown here.}
\label{fitresult}
\end{deluxetable}

\begin{figure*}[!htb]
\centering
\includegraphics[width=0.45\columnwidth]{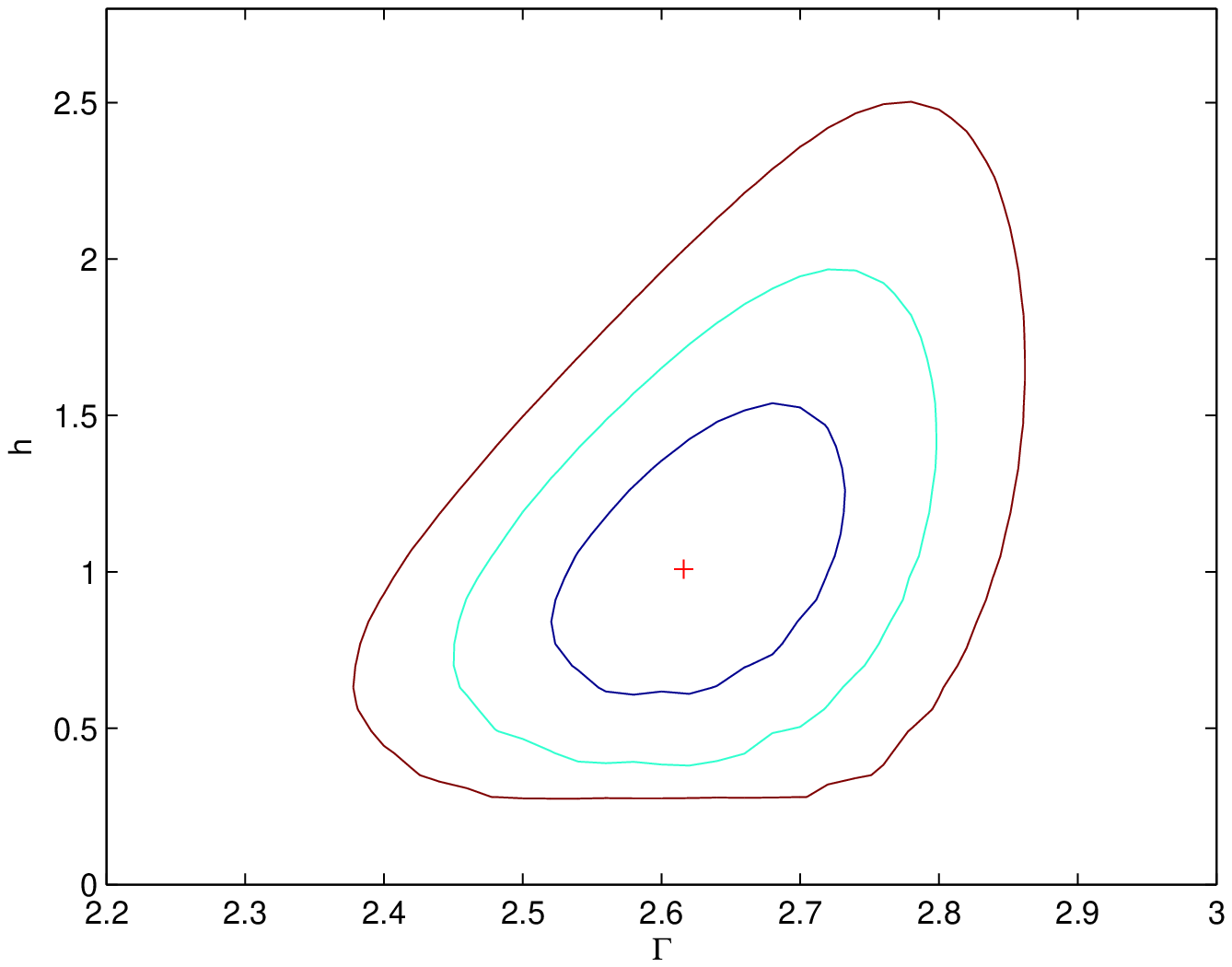}
\includegraphics[width=0.45\columnwidth]{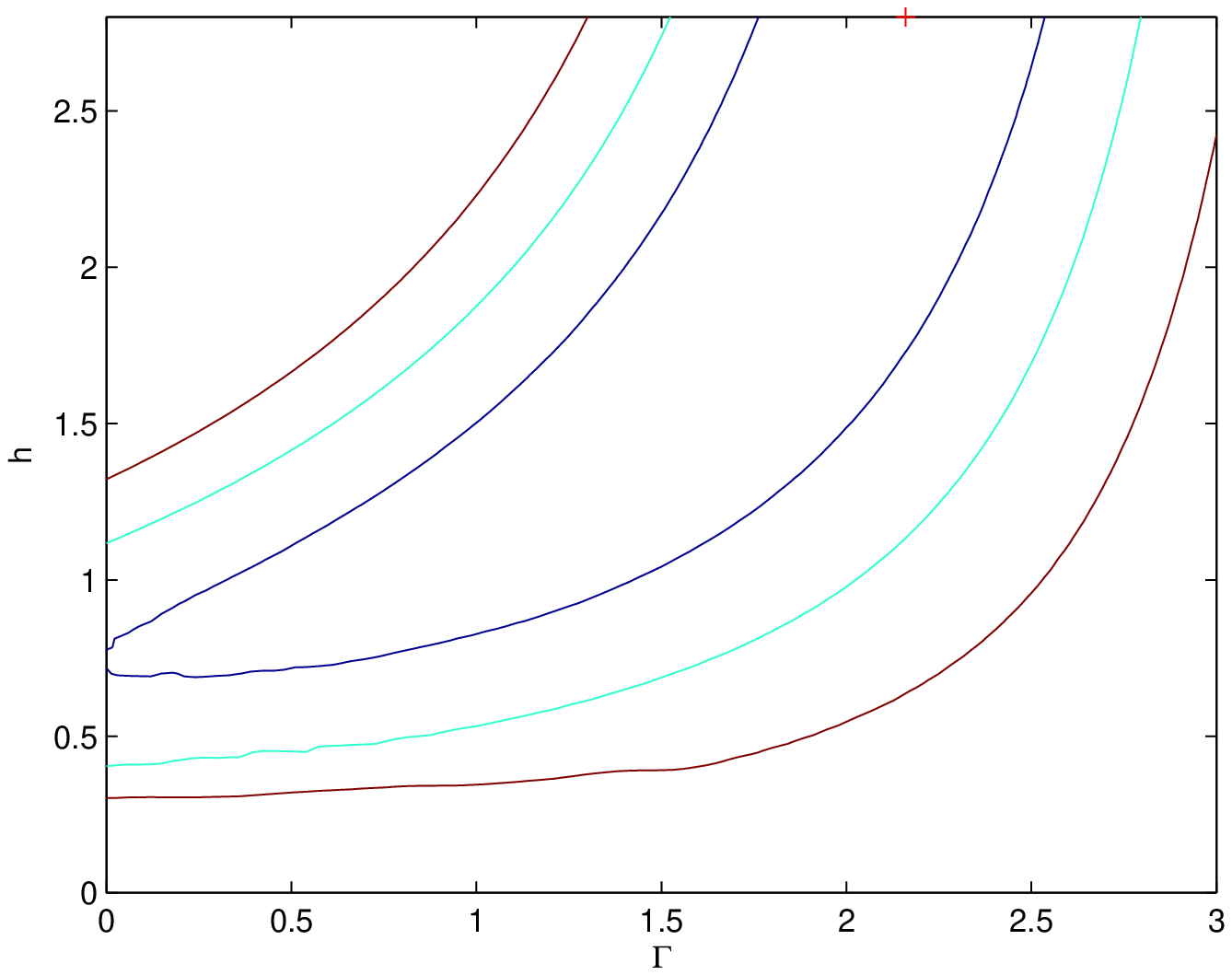}
\caption {Same as Fig. \ref{fixA} but with a prior $A=1\pm0.25$.} 
\label{Aprior}
\end{figure*}

Furthermore, we take the uncertainty of the CIB into account by employing 
a prior $A=1\pm0.25$ when doing the global fitting. The fitting results 
of $\Gamma_{\rm int}$ and $h$ are shown in Fig. \ref{Aprior}. For Mrk 501, 
the best-fitting values are almost the same as the case $A=1$ (Fig. 
\ref{fixA}), but the contours become larger after including the 
uncertainty of CIB. The fitted Hubble constant with $1\sigma$ range is 
$h=1.01^{+0.53}_{-0.40}$. While for 1ES 1101-232, the best-fitting 
values of the parameters differ significantly from the case $A=1$
and have larger uncertainties, as shown in the right panel of Fig. 
\ref{Aprior}. 

\begin{figure*}[!htb]
\centering
\includegraphics[width=0.45\columnwidth]{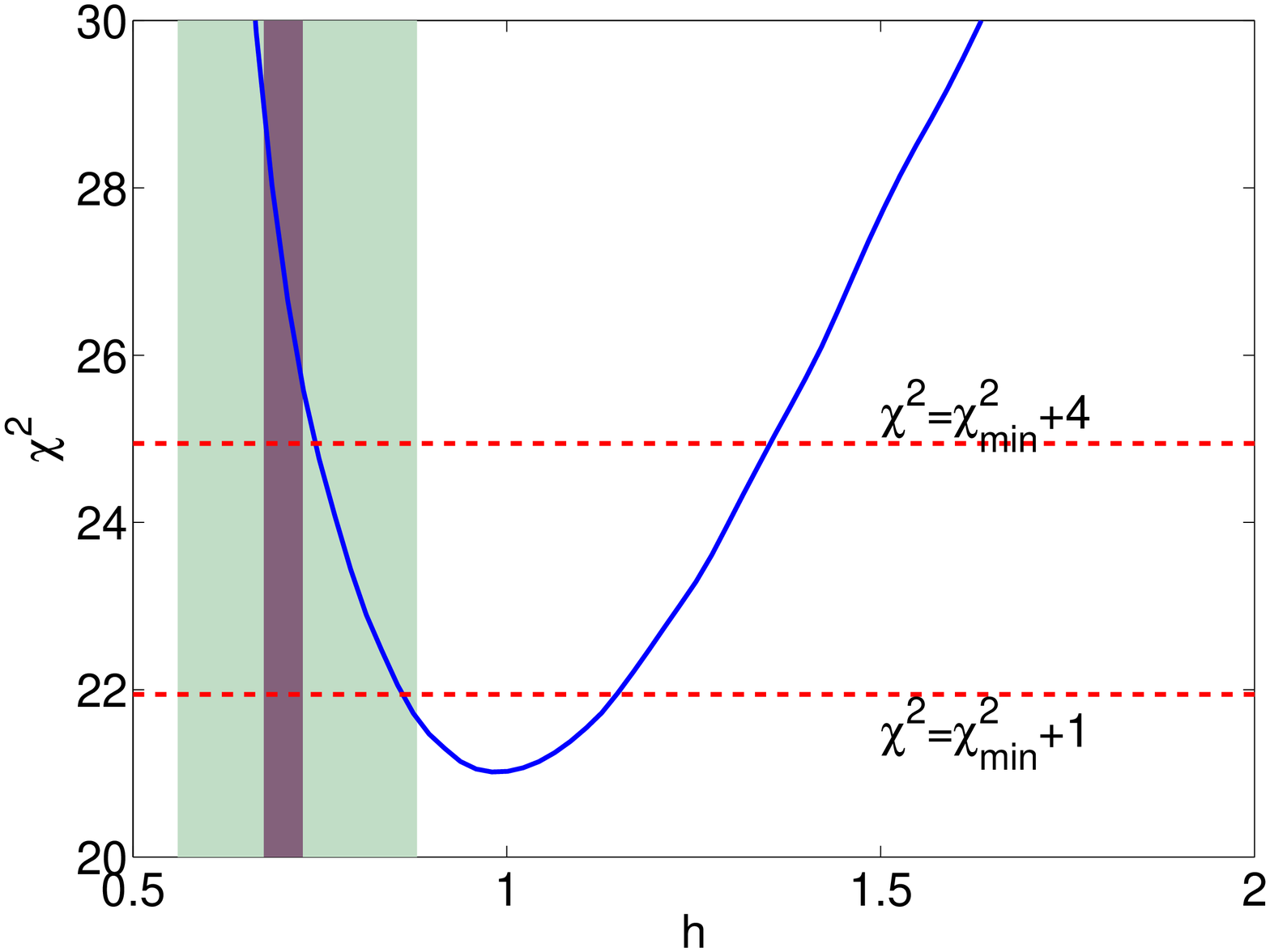}
\includegraphics[width=0.45\columnwidth]{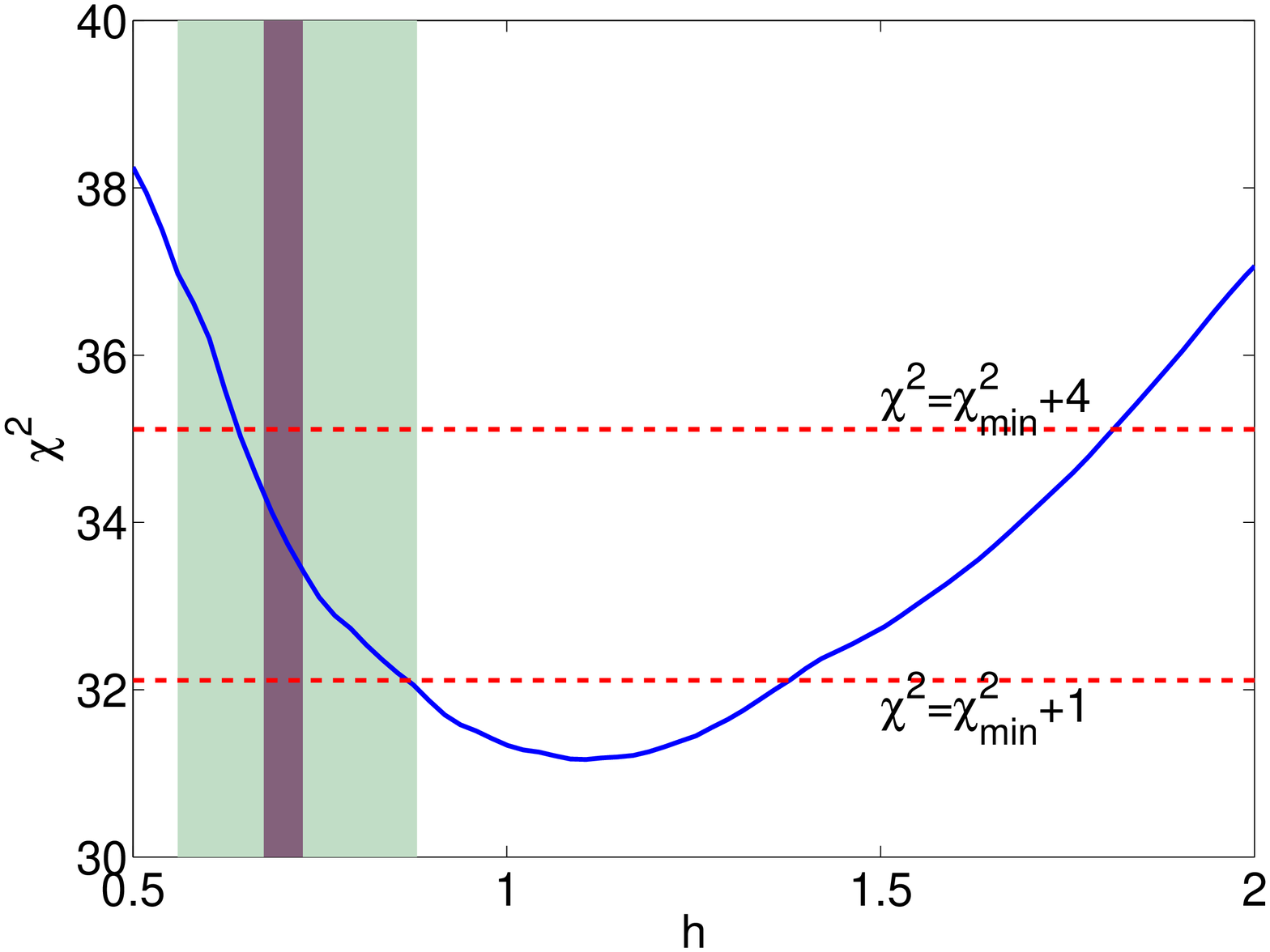}
\caption {$\chi^2$ as functions of $h$ for the combined fitting using both
Mrk 501 and 1ES 1101-232. Left: $A=1$; right: $A=1\pm0.25$. The two
horizon lines in each panel are $\chi^2=\chi^2_{\rm min}+1$ and
$\chi^2=\chi^2_{\rm min}+4$, which represent the $1\sigma$ and 
$2\sigma$ uncertainties of $h$. The shaded regions show the $2\sigma$
results on the Hubble constant $h=0.72\pm0.16$ (wider one) from the
Hubble Space Telescope (HST) Key Project \citep{2001ApJ...553...47F} 
and $h=0.701\pm 0.026$ (narrower one) from the recent observations of 
CMB, SNe Ia and baryon acoustic oscillation \citep{2008arXiv0803.0547K}.} 
\label{combine}
\end{figure*}

Finally, we combine data of the two sources to fit the Hubble constant $h$.
We show the $\chi^2$ values as functions of $h$ taking the prior $A=1$ 
and $A=1\pm0.25$ respectively in Fig. \ref{combine}. The lines $\chi^2=
\chi_{\rm min}^2+1$ ($\chi^2=\chi^2_{\rm min}+4$) is plotted to show 
the $1\sigma$ ($2\sigma$) range of parameter
$h$. We find that $h=1.00^{+0.15}_{-0.14}$ for $A=1$, $h=1.05^{+0.35}_
{-0.19}$ for $A=1\pm0.25$ at $1\sigma$ level respectively. We can see 
that after combining data of the two sources the best value of $h$ is 
not sensitive to the uncertainties of CIB, although the error bar of 
$h$ becomes larger.

\section{Conclusion and discussion}

In this work we constrain the cosmological parameters, 
especially the Hubble constant $H_0$, by observations of extragalactic 
VHE $\gamma$-ray sources at cosmological distances. The VHE $\gamma$-rays 
experience attenuation by background radiation field through $e^+e^-$ 
pair production. This attenuation is proportional to the distance that 
the VHE $\gamma$-rays go through. Therefore the absorption of VHE 
$\gamma$-rays can be used to determine the distance of VHE $\gamma$-ray 
sources, accordingly to get constraints on the cosmological parameters.

By fitting the spectra of two blazars Mrk 501 and 1ES 1101-232 we get 
the best-fitting Hubble constant is $H_0\sim 100$ km s$^{-1}$ Mpc$^{-1}$. 
A large Hubble constant implies that the absorption of VHE 
$\gamma$-rays is not as significant as we usually expected 
\citep{2000PhLB..493....1P,2006Natur.440.1018A}. Since the observations 
of VHE $\gamma$-rays and CIB are still rough, the errors of the 
fitting parameters are also very large. The mechanism constraining the 
Hubble constant adopted here is very different from previous methods, 
however, our results are consistent with the recent combined analysis 
of CMB, SNe Ia and baryon acoustic oscillation data on Hubble constant 
$h=0.701\pm0.026$ at $2\sigma$ level \citep{2008arXiv0803.0547K}.

In fact, for each single method measuring the Hubble constant there
are relatively large uncertainties, including both the statistical
and the systematic ones. The HST Key Project measured the Hubble 
constant from several secondary distance indicators
using Cepheid as calibration \citep{2001ApJ...553...47F}. They gave
the results that $h=0.71\pm0.06$ for SNe Ia, $h=0.71\pm0.08$ for the
Tully-Fisher relation of spiral galaxies, $h=0.70\pm0.08$ for the
surface brightness fluctuations of galaxies, $h=0.72\pm0.11$ for Type 
II supernovae and $h=0.82\pm0.11$ for the fundamental plane method of 
elliptical galaxies. The combined result of HST Key Project suggested
$h=0.72\pm0.08$. While after the CMB observations are involved, the
result is greatly improved \citep{2008arXiv0803.0547K}, as shown in Fig.
\ref{combine}. It shows that the cross check and combination of different 
methods are very helpful to find the right answer and improve the accuracy.

Before concluding we would like to briefly comment the simple assumptions 
adopted in our work. We assume the intrinsic spectrum to be a single 
power law. In fact the spectral energy distribution of many AGNs can 
be well described by the so-called synchrotron self-Compton (SSC) model. 
In the SSC scenario, the VHE $\gamma$-rays are produced through the 
inverse Compton (IC) scatterings between the electrons and synchrotron 
photons generated themselves. The VHE spectrum generally has an 
``IC peak'' originates from the transition from Thomson regime to 
Klein-Nishima regime \citep{1970RvMP...42..237B}. However, if the 
energy range is narrow, e.g., within a decade of energy, it can be 
described approximately by a power law. 

The CIB model and its uncertainties are also too simplified.
Because of the contamination of foreground radiation, it is usually 
difficult to get reliable CIB from measurements. The galaxy evolution 
models to predict CIB also have large uncertainties 
\citep{2001ARA&A..39..249H,2006ApJ...648..774S}. In 
\cite{2006ApJ...648..774S} the results between the ``baseline'' and 
``fast'' evolution models differ by about $20\% \sim 40\%$. While 
the results given by \cite{2005AIPC..745...23P} differ from 
\cite{2006ApJ...648..774S} by a factor of 2 at some wavelengths.

Anyway we think the present work is only a prototype of such studies,
since the observation of VHE $\gamma$ rays from extragalactic 
sources is achieved only in the recent years. This field is actually 
immature and at its early stage. With the next generation 
of space and ground-based instruments more extragalactic sources will 
be observed with high precision. Especially the space observatory
Fermi\footnote{See the homepage of Fermi, http://www-glast.stanford.edu} 
(energy range from 
MeV to hundred GeV) can explore $\gamma$-ray sources to redshift 
$z\sim 1$, due to an estimate of the ``$\gamma$-ray horizon'' 
$\log(z)\sim 1-0.7\log(E/1{\rm GeV})$ \citep{2007AIPC..921...24H}.
The ``$\gamma$-ray horizon'' is the redshift corresponding to absorption 
depth $\tau\approx 1$ for energy $E$. With larger sample of data, 
higher precision of spectra and higher redshift sources from Fermi we 
can even explore more cosmological parameters besides the Hubble constant, 
such as the cosmological component or the equation of state of dark energy. 
The development of ground-based instruments also aims to lower threshold 
energy and improve sensitivities. We anticipate the field of VHE 
$\gamma$-ray will develop quickly in the near future. Our work shows 
that the VHE $\gamma$-rays may become an more important field, not only 
to astrophysics but also to the cosmology.

\acknowledgments

This work is supported by the NSF of China under the grant
Nos. 10575111, 10773011 and supported in part by the Chinese Academy of
Sciences under the grant No. KJCX3-SYW-N2.

\bibliography{../../../cygnus/tex/gammaray}

\begin{thebibliography}{42}
\expandafter\ifx\csname natexlab\endcsname\relax\def\natexlab#1{#1}\fi

\bibitem[{{Aharonian} {et~al.}(1997){Aharonian}, {Akhperjanian}, {Barrio},
  {Bernloehr}, {Beteta}, {Bradbury}, {Contreras}, {Cortina}, {Daum}, {Deckers},
  {Feigl}, {Fernandez}, {Fonseca}, {Frass}, {Funk}, {Gonzalez}, {Haustein},
  {Heinzelmann}, {Hemberger}, {Hermann}, {Hess}, {Heusler}, {Hofmann}, {Holl},
  {Horns}, {Kankanian}, {Kirstein}, {Koehler}, {Konopelko}, {Kornmayer},
  {Kranich}, {Krawczynski}, {Lampeitl}, {Lindner}, {Lorenz}, {Magnussen},
  {Meyer}, {Mirzoyan}, {Moeller}, {Moralejo}, {Padilla}, {Panter}, {Petry},
  {Plaga}, {Prahl}, {Prosch}, {Puehlhofer}, {Rauterberg}, {Rhode}, {Rivero},
  {Roehring}, {Sahakian}, {Samorski}, {Sanchez}, {Schmele}, {Schmidt}, {Stamm},
  {Ulrich}, {Voelk}, {Westerhoff}, {Wiebel-Sooth}, {Wiedner}, {Willmer}, \&
  {Wirth}}]{1997A&A...327L...5A}
{Aharonian}, F., {Akhperjanian}, A.~G., {Barrio}, J.~A., {et~al.} 1997, \aap,
  327, L5

\bibitem[{{Aharonian} {et~al.}(2006){Aharonian}, {Akhperjanian}, {Bazer-Bachi},
  {Beilicke}, {Benbow}, {Berge}, {Bernl{\"o}hr}, {Boisson}, {Bolz}, {Borrel},
  {Braun}, {Breitling}, {Brown}, {Chadwick}, {Chounet}, {Cornils},
  {Costamante}, {Degrange}, {Dickinson}, {Djannati-Ata{\"i}}, {Drury}, {Dubus},
  {Emmanoulopoulos}, {Espigat}, {Feinstein}, {Fontaine}, {Fuchs}, {Funk},
  {Gallant}, {Giebels}, {Gillessen}, {Glicenstein}, {Goret}, {Hadjichristidis},
  {Hauser}, {Hauser}, {Heinzelmann}, {Henri}, {Hermann}, {Hinton}, {Hofmann},
  {Holleran}, {Horns}, {Jacholkowska}, {de Jager}, {Kh{\'e}lifi}, {Klages},
  {Komin}, {Konopelko}, {Latham}, {Le Gallou}, {Lemi{\`e}re},
  {Lemoine-Goumard}, {Leroy}, {Lohse}, {Martin}, {Martineau-Huynh},
  {Marcowith}, {Masterson}, {McComb}, {de Naurois}, {Nolan}, {Noutsos},
  {Orford}, {Osborne}, {Ouchrif}, {Panter}, {Pelletier}, {Pita},
  {P{\"u}hlhofer}, {Punch}, {Raubenheimer}, {Raue}, {Raux}, {Rayner}, {Reimer},
  {Reimer}, {Ripken}, {Rob}, {Rolland}, {Rowell}, {Sahakian}, {Saug{\'e}},
  {Schlenker}, {Schlickeiser}, {Schuster}, {Schwanke}, {Siewert}, {Sol},
  {Spangler}, {Steenkamp}, {Stegmann}, {Tavernet}, {Terrier}, {Th{\'e}oret},
  {Tluczykont}, {van Eldik}, {Vasileiadis}, {Venter}, {Vincent}, {V{\"o}lk}, \&
  {Wagner}}]{2006Natur.440.1018A}
{Aharonian}, F., {Akhperjanian}, A.~G., {Bazer-Bachi}, A.~R., {et~al.} 2006,
  \nat, 440, 1018

\bibitem[{{Aharonian}(2001)}]{2001ICRC...27I.250A}
{Aharonian}, F.~A. 2001, in International Cosmic Ray Conference, Vol.~27,
  International Cosmic Ray Conference, ed. G.~{Exarhos} \& X.~{Moussas}, I250

\bibitem[{{Aharonian} {et~al.}(2001){Aharonian}, {Akhperjanian}, {Barrio},
  {Bernl{\"o}hr}, {Bolz}, {B{\"o}rst}, {Bojahr}, {Contreras}, {Cortina},
  {Denninghoff}, {Fonseca}, {Gonzalez}, {G{\"o}tting}, {Heinzelmann},
  {Hermann}, {Heusler}, {Hofmann}, {Horns}, {Ibarra}, {Iserlohe}, {Jung},
  {Kankanyan}, {Kestel}, {Kettler}, {Kohnle}, {Konopelko}, {Kornmeyer},
  {Kranich}, {Krawczynski}, {Lampeitl}, {Lorenz}, {Lucarelli}, {Magnussen},
  {Mang}, {Meyer}, {Mirzoyan}, {Moralejo}, {Padilla}, {Panter}, {Plaga},
  {Plyasheshnikov}, {Prahl}, {P{\"u}hlhofer}, {Rhode}, {R{\"o}hring}, {Rowell},
  {Sahakian}, {Samorski}, {Schilling}, {Schr{\"o}der}, {Siems}, {Stamm},
  {Tluczykont}, {V{\"o}lk}, {Wiedner}, \& {Wittek}}]{2001A&A...366...62A}
{Aharonian}, F.~A., {Akhperjanian}, A.~G., {Barrio}, J.~A., {et~al.} 2001,
  \aap, 366, 62

\bibitem[{{Aharonian} {et~al.}(2008){Aharonian}, {Khangulyan}, \&
  {Costamante}}]{2008MNRAS.387.1206A}
{Aharonian}, F.~A., {Khangulyan}, D., \& {Costamante}, L. 2008, \mnras, 387,
  1206

\bibitem[{{Allen} {et~al.}(2008){Allen}, {Rapetti}, {Schmidt}, {Ebeling},
  {Morris}, \& {Fabian}}]{2008MNRAS.383..879A}
{Allen}, S.~W., {Rapetti}, D.~A., {Schmidt}, R.~W., {et~al.} 2008, \mnras, 383,
  879

\bibitem[{{Astier} {et~al.}(2006){Astier}, {Guy}, {Regnault}, {Pain},
  {Aubourg}, {Balam}, {Basa}, {Carlberg}, {Fabbro}, {Fouchez}, {Hook},
  {Howell}, {Lafoux}, {Neill}, {Palanque-Delabrouille}, {Perrett}, {Pritchet},
  {Rich}, {Sullivan}, {Taillet}, {Aldering}, {Antilogus}, {Arsenijevic},
  {Balland}, {Baumont}, {Bronder}, {Courtois}, {Ellis}, {Filiol}, {Gon{\c
  c}alves}, {Goobar}, {Guide}, {Hardin}, {Lusset}, {Lidman}, {McMahon},
  {Mouchet}, {Mourao}, {Perlmutter}, {Ripoche}, {Tao}, \&
  {Walton}}]{2006A&A...447...31A}
{Astier}, P., {Guy}, J., {Regnault}, N., {et~al.} 2006, \aap, 447, 31

\bibitem[{{Barrau} {et~al.}(2008){Barrau}, {Gorecki}, \&
  {Grain}}]{2008arXiv0804.3699B}
{Barrau}, A., {Gorecki}, A., \& {Grain}, J. 2008, ArXiv e-prints: 0804.3699

\bibitem[{{Blumenthal} \& {Gould}(1970)}]{1970RvMP...42..237B}
{Blumenthal}, G.~R. \& {Gould}, R.~J. 1970, Reviews of Modern Physics, 42, 237

\bibitem[{{Catanese} {et~al.}(1997){Catanese}, {Bradbury}, {Breslin},
  {Buckley}, {Carter-Lewis}, {Cawley}, {Dermer}, {Fegan}, {Finley}, {Gaidos},
  {Hillas}, {Johnson}, {Krennrich}, {Lamb}, {Lessard}, {Macomb}, {McEnery},
  {Moriarty}, {Quinn}, {Rodgers}, {Rose}, {Samuelson}, {Sembroski},
  {Srinivasan}, {Weekes}, \& {Zweerink}}]{1997ApJ...487L.143C}
{Catanese}, M., {Bradbury}, S.~M., {Breslin}, A.~C., {et~al.} 1997, \apjl, 487,
  L143

\bibitem[{{Coppi} \& {Aharonian}(1999)}]{1999APh....11...35C}
{Coppi}, P.~S. \& {Aharonian}, F.~A. 1999, Astroparticle Physics, 11, 35

\bibitem[{{de Bernardis} {et~al.}(2000){de Bernardis}, {Ade}, {Bock}, {Bond},
  {Borrill}, {Boscaleri}, {Coble}, {Crill}, {De Gasperis}, {Farese},
  {Ferreira}, {Ganga}, {Giacometti}, {Hivon}, {Hristov}, {Iacoangeli}, {Jaffe},
  {Lange}, {Martinis}, {Masi}, {Mason}, {Mauskopf}, {Melchiorri}, {Miglio},
  {Montroy}, {Netterfield}, {Pascale}, {Piacentini}, {Pogosyan}, {Prunet},
  {Rao}, {Romeo}, {Ruhl}, {Scaramuzzi}, {Sforna}, \&
  {Vittorio}}]{2000Natur.404..955D}
{de Bernardis}, P., {Ade}, P.~A.~R., {Bock}, J.~J., {et~al.} 2000, \nat, 404,
  955

\bibitem[{{Djannati-Atai } {et~al.}(1999){Djannati-Atai }, {Piron}, {Barrau},
  {Iacoucci}, {Punch}, {Tavernet}, {Bazer-Bachi}, {Cabot}, {Chounet},
  {Debiais}, {Degrange}, {Dezalay}, {Dumora}, {Espigat}, {Fabre}, {Fleury},
  {Fontaine}, {Ghesqui{\`e}re}, {Goret}, {Gouiffes}, {Grenier}, {Le Bohec},
  {Malet}, {Meynadier}, {Mohanty}, {Nuss}, {Par{\'e}}, {Qu{\'e}bert}, {Ragan},
  {Renault}, {Rivoal}, {Rob}, {Schahmaneche}, \& {Smith}}]{1999A&A...350...17D}
{Djannati-Atai }, A., {Piron}, F., {Barrau}, A., {et~al.} 1999, \aap, 350, 17

\bibitem[{{Freedman} {et~al.}(2001){Freedman}, {Madore}, {Gibson}, {Ferrarese},
  {Kelson}, {Sakai}, {Mould}, {Kennicutt}, {Ford}, {Graham}, {Huchra},
  {Hughes}, {Illingworth}, {Macri}, \& {Stetson}}]{2001ApJ...553...47F}
{Freedman}, W.~L., {Madore}, B.~F., {Gibson}, B.~K., {et~al.} 2001, \apj, 553,
  47

\bibitem[{{Gould} \& {Schr{\'e}der}(1966)}]{1966PhRvL..16..252G}
{Gould}, R.~J. \& {Schr{\'e}der}, G. 1966, Physical Review Letters, 16, 252

\bibitem[{{Gould} \& {Schr{\'e}der}(1967)}]{1967PhRv..155.1404G}
{Gould}, R.~J. \& {Schr{\'e}der}, G.~P. 1967, Physical Review, 155, 1404

\bibitem[{{Hartmann}(2007)}]{2007AIPC..921...24H}
{Hartmann}, D.~H. 2007, in American Institute of Physics Conference Series,
  Vol. 921, The First GLAST Symposium, ed. S.~{Ritz}, P.~{Michelson}, \& C.~A.
  {Meegan}, 24--25

\bibitem[{{Hauser} \& {Dwek}(2001)}]{2001ARA&A..39..249H}
{Hauser}, M.~G. \& {Dwek}, E. 2001, \araa, 39, 249

\bibitem[{{Hayashida} {et~al.}(1998){Hayashida}, {Hirasawa}, {Ishikawa},
  {Lafoux}, {Nagano}, {Nishikawa}, {Ouchi}, {Ohoka}, {Ohnishi}, {Sakaki},
  {Sasaki}, {Shimodaira}, {Teshima}, {Torii}, {Yamamoto}, {Yoshida}, {Yuda},
  {Hayashi}, {Ito}, {Kawasaki}, {Kawasaki}, {Matsuyama}, {Sasano}, {Takahashi},
  {Chamoto}, {Kajino}, {Sakata}, {Sugiyama}, {Tsukiji}, {Yamamoto}, {Inoue},
  {Kusano}, {Mizutani}, {Shiomi}, {Hibino}, {Kashiwagi}, {Nishimura}, {Loh},
  {Sokolsky}, {Taylor}, {Honda}, {Kawasumi}, {Tsushima}, {Uchihori},
  {Kitamura}, {Chikawa}, {Kabe}, {Mizumoto}, {Yoshii}, {Hotta}, {Saito},
  {Nishizawa}, {Kuramochi}, \& {Sakumoto}}]{1998ApJ...504L..71H}
{Hayashida}, N., {Hirasawa}, H., {Ishikawa}, F., {et~al.} 1998, \apjl, 504, L71

\bibitem[{{Hooper} \& {Serpico}(2007)}]{2007PhRvL..99w1102H}
{Hooper}, D. \& {Serpico}, P.~D. 2007, Physical Review Letters, 99, 231102

\bibitem[{{Komatsu} {et~al.}(2008){Komatsu}, {Dunkley}, {Nolta}, {Bennett},
  {Gold}, {Hinshaw}, {Jarosik}, {Larson}, {Limon}, {Page}, {Spergel},
  {Halpern}, {Hill}, {Kogut}, {Meyer}, {Tucker}, {Weiland}, {Wollack}, \&
  {Wright}}]{2008arXiv0803.0547K}
{Komatsu}, E., {Dunkley}, J., {Nolta}, M.~R., {et~al.} 2008, ArXiv e-prints:
  0803.0547

\bibitem[{{Krawczynski} {et~al.}(2000){Krawczynski}, {Coppi}, {Maccarone}, \&
  {Aharonian}}]{2000A&A...353...97K}
{Krawczynski}, H., {Coppi}, P.~S., {Maccarone}, T., \& {Aharonian}, F.~A. 2000,
  \aap, 353, 97

\bibitem[{{Malkov} \& {O'C Drury}(2001)}]{2001RPPh...64..429M}
{Malkov}, M.~A. \& {O'C Drury}, L. 2001, Reports of Progress in Physics, 64,
  429

\bibitem[{{Munshi} {et~al.}(2008){Munshi}, {Valageas}, {van Waerbeke}, \&
  {Heavens}}]{2008PhR...462...67M}
{Munshi}, D., {Valageas}, P., {van Waerbeke}, L., \& {Heavens}, A. 2008,
  \physrep, 462, 67

\bibitem[{{Nikishov}(1962)}]{Nikishov1962}
{Nikishov}, A.~I. 1962, Soviet Physics JEPT, 14, 2

\bibitem[{{Perlmutter} {et~al.}(1999){Perlmutter}, {Aldering}, {Goldhaber},
  {Knop}, {Nugent}, {Castro}, {Deustua}, {Fabbro}, {Goobar}, {Groom}, {Hook},
  {Kim}, {Kim}, {Lee}, {Nunes}, {Pain}, {Pennypacker}, {Quimby}, {Lidman},
  {Ellis}, {Irwin}, {McMahon}, {Ruiz-Lapuente}, {Walton}, {Schaefer}, {Boyle},
  {Filippenko}, {Matheson}, {Fruchter}, {Panagia}, {Newberg}, {Couch}, \& {The
  Supernova Cosmology Project}}]{1999ApJ...517..565P}
{Perlmutter}, S., {Aldering}, G., {Goldhaber}, G., {et~al.} 1999, \apj, 517,
  565

\bibitem[{{Primack} {et~al.}(2005){Primack}, {Bullock}, \&
  {Somerville}}]{2005AIPC..745...23P}
{Primack}, J.~R., {Bullock}, J.~S., \& {Somerville}, R.~S. 2005, in American
  Institute of Physics Conference Series, Vol. 745, High Energy Gamma-Ray
  Astronomy, ed. F.~A. {Aharonian}, H.~J. {V{\"o}lk}, \& D.~{Horns}, 23--33

\bibitem[{{Primack} {et~al.}(1999){Primack}, {Bullock}, {Somerville}, \&
  {MacMinn}}]{1999APh....11...93P}
{Primack}, J.~R., {Bullock}, J.~S., {Somerville}, R.~S., \& {MacMinn}, D. 1999,
  Astroparticle Physics, 11, 93

\bibitem[{{Protheroe} \& {Meyer}(2000)}]{2000PhLB..493....1P}
{Protheroe}, R.~J. \& {Meyer}, H. 2000, Physics Letters B, 493, 1

\bibitem[{{Punch} {et~al.}(1992){Punch}, {Akerlof}, {Cawley}, {Chantell},
  {Fegan}, {Fennell}, {Gaidos}, {Hagan}, {Hillas}, {Jiang}, {Kerrick}, {Lamb},
  {Lawrence}, {Lewis}, {Meyer}, {Mohanty}, {O'Flaherty}, {Reynolds}, {Rovero},
  {Schubnell}, {Sembroski}, {Weekes}, \& {Wilson}}]{1992Natur.358..477P}
{Punch}, M., {Akerlof}, C.~W., {Cawley}, M.~F., {et~al.} 1992, \nat, 358, 477

\bibitem[{{Quinn} {et~al.}(1996){Quinn}, {Akerlof}, {Biller}, {Buckley},
  {Carter-Lewis}, {Cawley}, {Catanese}, {Connaughton}, {Fegan}, {Finley},
  {Gaidos}, {Hillas}, {Lamb}, {Krennrich}, {Lessard}, {McEnery}, {Meyer},
  {Mohanty}, {Rodgers}, {Rose}, {Sembroski}, {Schubnell}, {Weekes}, {Wilson},
  \& {Zweerink}}]{1996ApJ...456L..83Q}
{Quinn}, J., {Akerlof}, C.~W., {Biller}, S., {et~al.} 1996, \apjl, 456, L83

\bibitem[{{Renault} {et~al.}(2001){Renault}, {Barrau}, {Lagache}, \&
  {Puget}}]{2001A&A...371..771R}
{Renault}, C., {Barrau}, A., {Lagache}, G., \& {Puget}, J.-L. 2001, \aap, 371,
  771

\bibitem[{{Riess} {et~al.}(1998){Riess}, {Filippenko}, {Challis},
  {Clocchiatti}, {Diercks}, {Garnavich}, {Gilliland}, {Hogan}, {Jha},
  {Kirshner}, {Leibundgut}, {Phillips}, {Reiss}, {Schmidt}, {Schommer},
  {Smith}, {Spyromilio}, {Stubbs}, {Suntzeff}, \&
  {Tonry}}]{1998AJ....116.1009R}
{Riess}, A.~G., {Filippenko}, A.~V., {Challis}, P., {et~al.} 1998, \aj, 116,
  1009

\bibitem[{{Riess} {et~al.}(2004){Riess}, {Strolger}, {Tonry}, {Casertano},
  {Ferguson}, {Mobasher}, {Challis}, {Filippenko}, {Jha}, {Li}, {Chornock},
  {Kirshner}, {Leibundgut}, {Dickinson}, {Livio}, {Giavalisco}, {Steidel},
  {Ben{\'{\i}}tez}, \& {Tsvetanov}}]{2004ApJ...607..665R}
{Riess}, A.~G., {Strolger}, L.-G., {Tonry}, J., {et~al.} 2004, \apj, 607, 665

\bibitem[{{Salamon} {et~al.}(1994){Salamon}, {Stecker}, \& {de
  Jager}}]{1994ApJ...423L...1S}
{Salamon}, M.~H., {Stecker}, F.~W., \& {de Jager}, O.~C. 1994, \apjl, 423, L1

\bibitem[{{Samuelson} {et~al.}(1998){Samuelson}, {Biller}, {Bond}, {Boyle},
  {Bradbury}, {Breslin}, {Buckley}, {Burdett}, {Buss'ons Gordo},
  {Carter-Lewis}, {Cantanese}, {Cawley}, {Fegan}, {Finley}, {Gaidos}, {Hall},
  {Hillas}, {Krennrich}, {Lamb}, {Lessard}, {McEnery}, {Masterson}, {Quinn},
  {Rodgers}, {Rose}, {Sembroski}, {Srinivasan}, {Vassiliev}, {Weekes}, \&
  {Zweerink}}]{1998ApJ...501L..17S}
{Samuelson}, F.~W., {Biller}, S.~D., {Bond}, I.~H., {et~al.} 1998, \apjl, 501,
  L17

\bibitem[{{Simet} {et~al.}(2008){Simet}, {Hooper}, \&
  {Serpico}}]{2008PhRvD..77f3001S}
{Simet}, M., {Hooper}, D., \& {Serpico}, P.~D. 2008, \prd, 77, 063001

\bibitem[{{Simon} {et~al.}(2005){Simon}, {Verde}, \&
  {Jimenez}}]{2005PhRvD..71l3001S}
{Simon}, J., {Verde}, L., \& {Jimenez}, R. 2005, \prd, 71, 123001

\bibitem[{{Spergel} {et~al.}(2003){Spergel}, {Verde}, {Peiris}, {Komatsu},
  {Nolta}, {Bennett}, {Halpern}, {Hinshaw}, {Jarosik}, {Kogut}, {Limon},
  {Meyer}, {Page}, {Tucker}, {Weiland}, {Wollack}, \&
  {Wright}}]{2003ApJS..148..175S}
{Spergel}, D.~N., {Verde}, L., {Peiris}, H.~V., {et~al.} 2003, \apjs, 148, 175

\bibitem[{{Stecker} {et~al.}(1992){Stecker}, {de Jager}, \&
  {Salamon}}]{1992ApJ...390L..49S}
{Stecker}, F.~W., {de Jager}, O.~C., \& {Salamon}, M.~H. 1992, \apjl, 390, L49

\bibitem[{{Stecker} {et~al.}(2006){Stecker}, {Malkan}, \&
  {Scully}}]{2006ApJ...648..774S}
{Stecker}, F.~W., {Malkan}, M.~A., \& {Scully}, S.~T. 2006, \apj, 648, 774

\bibitem[{{Tegmark} {et~al.}(2004){Tegmark}, {Blanton}, {Strauss}, {Hoyle},
  {Schlegel}, {Scoccimarro}, {Vogeley}, {Weinberg}, {Zehavi}, {Berlind},
  {Budavari}, {Connolly}, {Eisenstein}, {Finkbeiner}, {Frieman}, {Gunn},
  {Hamilton}, {Hui}, {Jain}, {Johnston}, {Kent}, {Lin}, {Nakajima}, {Nichol},
  {Ostriker}, {Pope}, {Scranton}, {Seljak}, {Sheth}, {Stebbins}, {Szalay},
  {Szapudi}, {Verde}, {Xu}, {Annis}, {Bahcall}, {Brinkmann}, {Burles},
  {Castander}, {Csabai}, {Loveday}, {Doi}, {Fukugita}, {Gott}, {Hennessy},
  {Hogg}, {Ivezi{\'c}}, {Knapp}, {Lamb}, {Lee}, {Lupton}, {McKay}, {Kunszt},
  {Munn}, {O'Connell}, {Peoples}, {Pier}, {Richmond}, {Rockosi}, {Schneider},
  {Stoughton}, {Tucker}, {Vanden Berk}, {Yanny}, \&
  {York}}]{2004ApJ...606..702T}
{Tegmark}, M., {Blanton}, M.~R., {Strauss}, M.~A., {et~al.} 2004, \apj, 606,
  702

\end{thebibliography}
\bibliographystyle{aa}

\end{document}